**Title: Chiroptical Nonlinearity Found in Covalent Organic Frameworks**

**Authors**: Yile Zhang[1], Lei Wei[1], Sixue Tang[1], Yuhang Li[1], Yanhang Ma*[1,3], Yue-Biao Zhang*[1,3,4], and Xiaoqin Shen*[1,2]

**Affiliations:**

[1]School of Physical Science and Technology, ShanghaiTech University, Shanghai, 201210, China.

[2]State Key Laboratory of Quantum Functional Materials, ShanghaiTech University, Shanghai, 201210, China.

[3]Shanghai Key Laboratory of High-Resolution Electron Microscopy, ShanghaiTech University, Shanghai 201210, China.

[4]Institute of Carbon Neutrality, State Key Laboratory of Advanced Medical Materials and Devices, ShanghaiTech University, Shanghai 201210, China.

*Corresponding authors. Email: shenxq@shanghaitech.edu.cn, zhangyb@shanghaitech.edu.cn, and mayh2@shanghaitech.edu.cn

**Abstract:** Chiroptical nonlinearity is vital for modern optics and photonics. However, it remains a central challenge to obtain a material simultaneously with large second-order nonlinearity and strong nonlinear optical chirality. Here we uncover the strong chiroptical nonlinearity in COF-330-a, owing to the formation of unique chiral, polar structure with large π-electron conjugation via structural transformation. A large effective second-order nonlinear susceptibility ($\chi_{\text{eff}}^{(2)} \approx 101$ pm/V) and a strong nonlinear chiroptical response ($g_{\text{SHG-CD}}$: -1.60 to +1.66) have been demonstrated in individual COF-330-a microcrystals, transcending the performance trade-off limitation in conventional chiroptical materials. These findings open a new avenue for designing nonlinear chiroptical materials which harness the duality of polarity and chirality and chemical versatility of COFs.

**Main Text:**

Obtaining a medium with strong chiroptical nonlinearity is highly desirable for various advanced applications in both classic and quantum nonlinear optics (*1*-*3*), such as optical encryption, holographic imaging and quantum information processing (*4*-*6*). In such a medium, second-order nonlinear optical processes would response distinctly under different circularly polarized fundamental waves (*7*), enabling direct control of light polarization in the nonlinear optical processes. Yet, its performance has been hindered by the fundamental dilemma: chiral crystals generally lack strong second-order nonlinearity, while conventional nonlinear optical crystals with strong polar structures tend to suppress chirality. To overcome this challenge, extensive efforts have been devoted to the exploration of nonlinear chiroptical materials. However, existing approaches are largely confined to incorporating chiral elements into inorganic crystals (*8*-*10*), often with an inherent performance trade-off between chirality and nonlinearity.

Reticular chemistry, which enables the construction of extended two- and three-dimensional organic porous crystals with defined topologies, offers robust platforms for developing functional materials (*11*-*13*). Particularly, covalent organic frameworks (COFs) exhibit exclusive advantages including high thermostability and extended π-conjugation systems, which are beneficial for tuning optical properties (*14*, *15*). One notable feature of three-dimensional COFs is their dynamic structural transformation between two or more single crystalline phases with identical topology. While 3D COFs are typical synthesized by using centrosymmetric organic monomers, structural transformation can lead to non-centrosymmetric crystal structures (*16-19*). Yet, the underlying symmetry-breaking mechanism and their potentials in second-order nonlinear optics have not yet been investigated.

Here we discover the coexistence of polar and chiral structures in non-centrosymmetric COF-330-a and demonstrate the exceptional chiroptical nonlinearity. The polar and chiral structural duality is induced by tetrahedral node rotation and linker twisting via structural transformation. We reveal a high second-order nonlinear susceptibility ($\chi^{(2)}_{\mathrm{eff}}$ ~ 101 pm/V) and a large nonlinear chiroptical activities (SHG circular dichroism: $|g_{\mathrm{SHG-CD}}| \approx 1.66$) in COF-330-a microcrystals, which transcends the performance trade-off limitation in other chiroptical materials. These findings open a new avenue for COFs in emerging photonic technologies such as chiral nonlinear optics and quantum photonics.

## Formation of polar and chiral structure in COF-330-a

COF-330 was synthesized from achiral monomers with acceptor-donor-acceptor (A-D-A) extended π-conjugation and adopted a ten-fold interpenetrated **dia** topology (*17*). When the guest molecules were gradually removed from pores, COF-330 experienced crystal structural transformations from a centrosymmetric space group $P4_2/n$ (COF-330-o) to another centrosymmetric space group $P2/n$ (COF-330-i) and then to a polar space group $Pnn2$ (COF-330-a) (*17*) (Fig. 1A and Fig. S1). Analysis of the skeleton geometries revealed that, COF-330-a underwent a self-induction chirality formation process in the transition from centrosymmetric to non-centrosymmetric structure, whereby the achiral linkers self-resolved into chiral, polar conformers of opposite handedness (Fig. 1B and Fig. S2). In this process, the central and vertex tetrahedral nodes within a unit cell rotated clockwise and counterclockwise of ~39°, respectively, about the *c* axis (Fig. S1), resulting in distinct asymmetric twisting of the linkers. The four linkers tethered to a central tetrahedral node of a repeating unit adopted pronounced left- and right-handed twists, two of each type, giving rise to two pair twisted chiral, polar conformers of opposite handedness (Fig. 1C). With this unique structural arrangement, slight un-equilibrium of the conformational chirality in the twisted linker pairs within a repeating unit would ultimately lead to a net chirality in the polar framework.

## Strong second-order nonlinearity in COF-330-a

The second-order nonlinear properties of COF-330-a microcrystals were investigated by using single-particle SHG micro-spectroscopy under a back-reflection configuration, with scanning electron microscopy (SEM) imaging for single-particle identification. (Fig. S3, Supplemental Text S2). COF-330-a microcrystals (~1.5 μm long, ~0.5 μm wide) have a slightly deformed tetragonal prism morphology (Fig. 2A and Fig 3). Under excitation with a femtosecond laser, individual COF-330-a generated intense blue SHG emission observed through the optical microscope (Fig. 2B). The SHG intensity gradually increased when approaching a longer wavelength and reached the maximum at ~ 445 nm **(**Fig. 2C**)**, which is consistent with the experimental optical absorption spectrum (Fig. S4), indicating considerable resonance enhancement. The SHG intensity shows a quadratic dependence on the excitation power (Fig. 2D) with a high laser-induced damage threshold of ~ 90 $GW \cdot cm^{-2}$ of individual crystals (Fig. S5). Polarization-dependent SHG measurement showed a two-lobed intensity pattern (Fig. 2E and Fig.

S6), with the maximum aligned with the microcrystal's longitudinal direction, indicating the polar crystalline *c* axis. Figure 2F shows a direct comparison of the SHG signals from an individual COF-330-a microcrystal and a reference microcrystal of urea. Remarkably, the $\chi^{(2)}_{\mathrm{eff}}$ of COF-330 is calculated to be ~101 pm/V (Supplementary Text S3), exceeding that of common inorganic benchmarks (e.g. $LiNbO_3$) (*20*) and SHG-active MOFs (*21, 22*).

The generation SHG activity was further investigated via in-situ single-particle SHG measurements, and the structural correlation was validated by three-dimensional electron diffraction (3D ED) (Fig. 3A and Fig. S7). At room temperature and atmospheric pressure, exclusion of guest ethanol molecules from the channels triggered the generation of SHG signal in COF-330. Upon exposure to a low partial pressure of guest vapor flow, the in-situ SHG signal of a COF-330-a microcrystal gradually diminished and ultimately vanished (Fig. 3B), along with gradual arising of faint two-photon excited fluorescence. The SHG signal can be recovered by controlling the guest atmosphere (Fig. 3C). The microcrystal lacking SHG activity adopted the *P*2/*n* space group, corresponding to the intermediate COF-330-i phase (Fig. 3, D and E). In contrast, the activated microcrystal with intense SHG activity adopted the *Pnn*2 space group (Fig. 3, F and G), consistent with the activated COF-330-a phase. The activation of second-order nonlinearity was further understood by density functional theory calculation (Fig. S8 and Supplementary text S3), which revealed that the asymmetric twisting of the large π-electron conjugated linkers generates a substantial electronic dipole moment of ~ 4.9 D per linker and a large molecular hyperpolarizability of ~ $62 \times 10^{-30}$ esu per linker in COF-330-a.

**Pronounced nonlinear chiroptical response in individual COF-330-a**

The prominent SHG activity of COF-330-a allows us to probe its net chirality at individual microcrystal level by using SHG circular dichroism (SHG-CD). Individual microcrystals were excited with left- and right-circularly polarized (LCP and RCP) femtosecond laser (Fig. 4A) and the degree of SHG-CD was quantified by the SHG-CD dissymmetry factor, defined as $g_{\mathrm{SHG-CD}} = 2(I_L-I_R)/(I_L+I_R)$, where $I_L$ and $I_R$ are the SHG intensities under LCP and RCP excitation, respectively. As shown in Fig. 4, B and C, the SHG intensities of two representative individual microcrystals depended strongly on the excitation helicity and showed opposite handedness responses, indicating the opposite net handedness of two microcrystals. The $g_{\mathrm{SHG-CD}}$ values were calculated to be +1.64 and -1.53, respectively. Statistical analysis of signals from 80 individual microcrystals showed the $g_{\mathrm{SHG-CD}}$ values spanned a wide range from –1.6 to +1.66, with an overall nearly equal population distribution of two opposite chiroptical responses (Fig. 4D). These results indicate that the local chiral distortion of twisted linkers of opposite handedness do not mutually compensate to each other within a microcrystal, yielding a net chirality in the polar framework.

A performance comparison with state-of-the-art materials demonstrates the unique advance of COF-330-a (Fig. 4E). A persuasive bottleneck in chiral nonlinear optical materials lies in an inherent performance trade-off: high $|g_{\mathrm{SHG-CD}}|$ typically coincides with small $\chi^{(2)}_{\mathrm{eff}}$, and vice versa. The material platforms include perovskites (*23-26*), metal complexes (*26-29*), coordinated polymers (*30*), and gold nanostructures (*31*). One breakthrough is chiral stacking of multiwall boron nitride nanotubes, which can achieve a large $|g_{\mathrm{SHG}}|$ of 1.4 and a moderate $\chi^{(2)}_{\mathrm{eff}}$ of 47 pm/V (*9*), yet further improving its nonlinearity may become challenging. In distinct contrast, COF-330-a transcends the performance trade-off limitation, achieving a high $|g_{\mathrm{SHG-CD}}|$ of 1.66 and a substantially large $\chi^{(2)}_{\mathrm{eff}}$ of 101 pm/V.

## Conclusion

We have revealed strong chiroptical nonlinearity in COF-330-a, which exhibits unique polar and chiral structures induced by a self-resolved structural transformation upon activation. A high second-order nonlinear susceptibility ($\chi_{\text{eff}}^{(2)}$ ~ 101 pm/V) and a simultaneous large nonlinear chiroptical activities ($g_{\text{SHG-CD}}$: from -1.60 to +1.66) can be achieved, endowed by the chiral twisted linkers with large π-electron conjugation and their polar arrangement. This performance constitutes a state-of-the-art advance that overcomes the long-standing trade-off limitation between second-order nonlinearity and chirality in conventional organic molecules and crystals (*32*). The materials also exhibit a high laser damage threshold (~ 90 GW/cm$^2$), showing the excellent stability. Given the chemical versatility and structural duality, the chiroptical nonlinearity of COFs can be further optimized and functionalized by rational molecular design. Our discovery establishes COFs as a versatile material platform that transcends traditional performance limitations in chiral nonlinear optics (*4*), opening new avenues for chiral quantum light sources (*33*, *34*), photon modulators and encryption (*35, 36*).

**Acknowledgments:**

**Funding:** This project was supported by the National Natural Science Foundation of China (Grant No. 62275152 and 12574362), the Science and Technology Commission of Shanghai Municipality (YDZX20253100001002). The authors also thank C*ħ*EM SPST, ShanghaiTech University (No. EM02161943), for characterization support.

**Author contributions:** XQS conceived the idea and led the project. YLZ conducted the nonlinear optical investigation. LW synthesized the COF samples. YLZ, LW and YHL conducted the SEM and 3D ED. SXT conducted the DFT calculation. LW, YHM, YLZ, and XQS discussed the results. X.S. interpreted the results and wrote the manuscript.

**Competing interests:** Authors declare that they have no competing interests.

**Data, code, and materials availability:** Crystallographic data reported in this paper are achieved from related references or the Cambridge Crystallographic Data Center (CCDC). All other data needed to evaluate to conclusions are available in the main text or the supplementary materials.

**Supplementary Materials**

Materials and Methods

Supplementary Text S1 to S6

Figs. S1 to S8

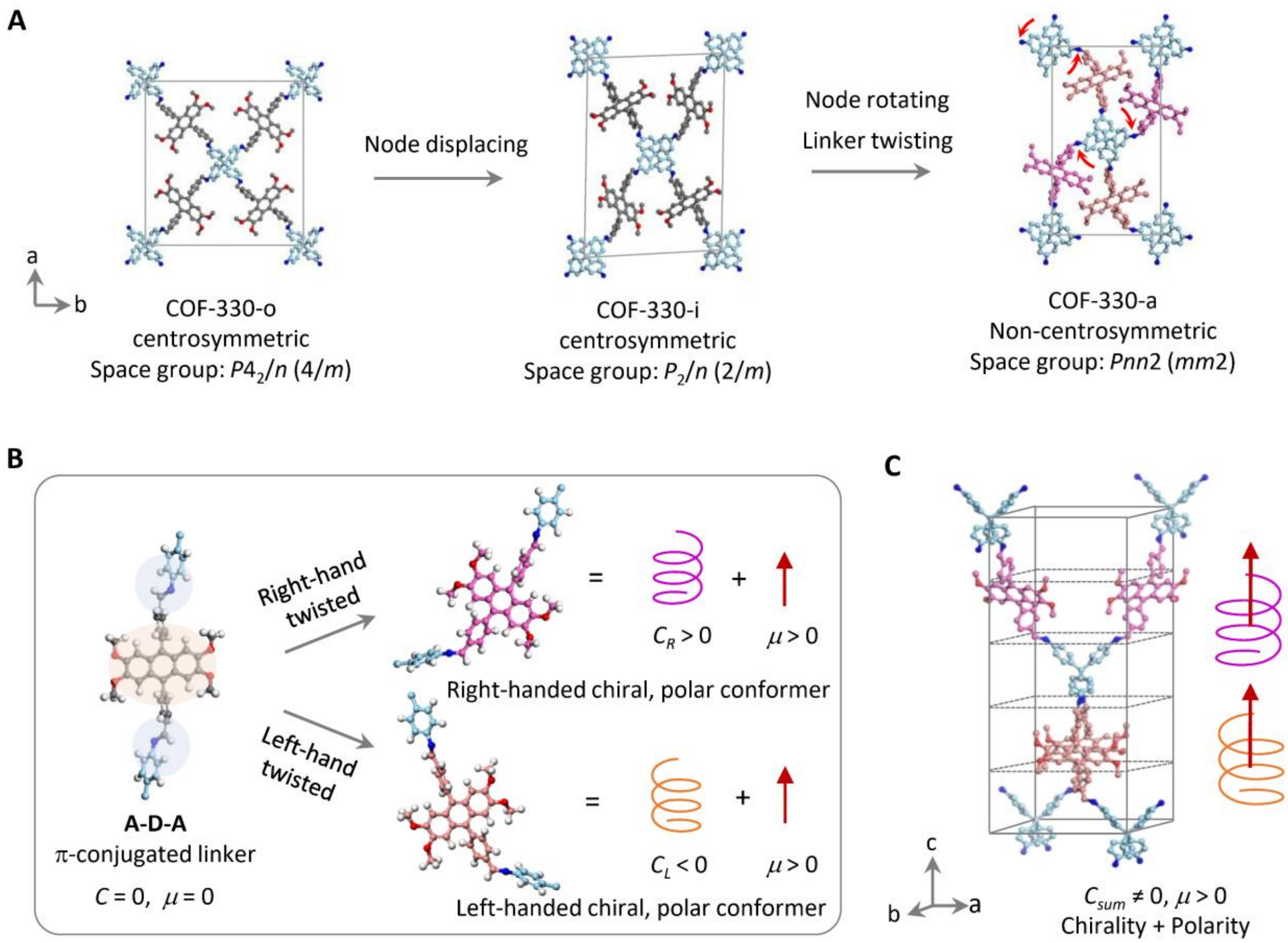

**Fig. 1. Formation of chirality and polarity in COF-330-a. (A)** Structural transformation from COF-330-o COF-330-a, involving node displacing in the first step and node rotating in the second step along with linker twisting. The right- and left-hand twisted linkers are encoded in pink and orange colors, respectively. The guest molecules were omitted for clarity. **(B)** Illustration of self-resolved chiral formation of the acceptor-donor-acceptor (A-D-A) π-conjugated linker, leading to a pair of chiral, polar conformers of opposite handedness ($C_R > 0$, $C_L < 0$). **(C)** A four-armed repeating unit across five-stacked unit cells for displaying the chiral and polar substructures of COF-330-a. Un-equilibrium of the opposite conformational chirality leads to a net chirality in the polar framework.

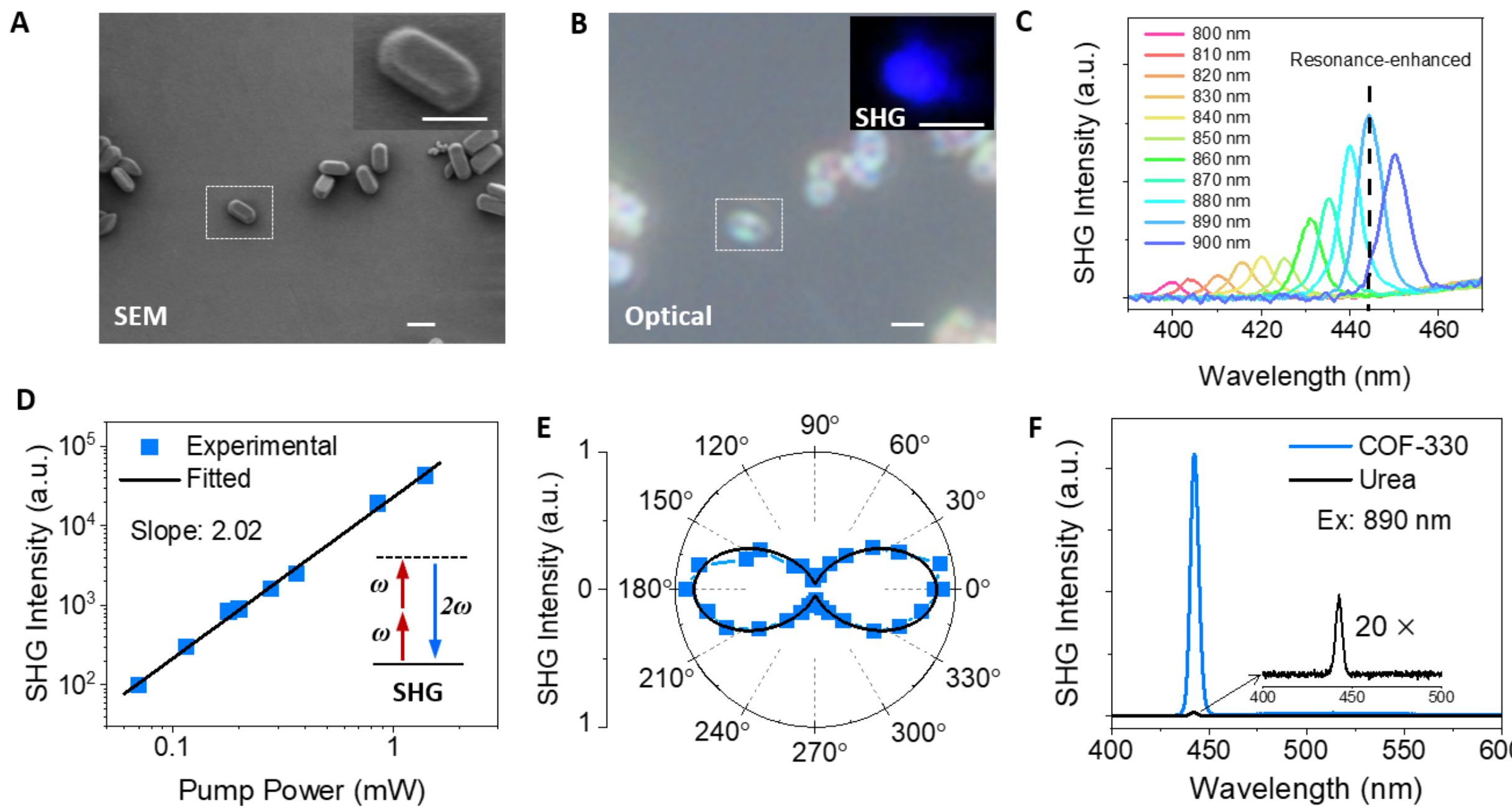

**Fig. 2. Strong SHG activity in individual COF-330. (A)** A representative SEM image of the COF-330 microcrystals for single-particle colocation. Inset is a close view of an individual crystal in the dashed rectangle. (**B**) An optical image of the same sample corresponding the SEM image in (A). Inset is a SHG image from the individual crystal in the dashed rectangle. (**C**) A representative SHG spectra of an individual crystal under exaction at wavelengths from 800 nm to 900 nm. (**D**) Pump-power-dependent SHG intensity. A slope of ~2 is consistent with the quadratic power dependency of SHG process. Inset is the schematic of SHG process. (**E**) Polarization-dependent SHG, indicating a highly anisotropic SHG response. The polarization angle along the longitudinal direction of the crystal was set as 0°. (**F**) Representative SHG spectra of an individual COF-330 microcrystal and a urea crystal reference. Scale bar: 1μm.

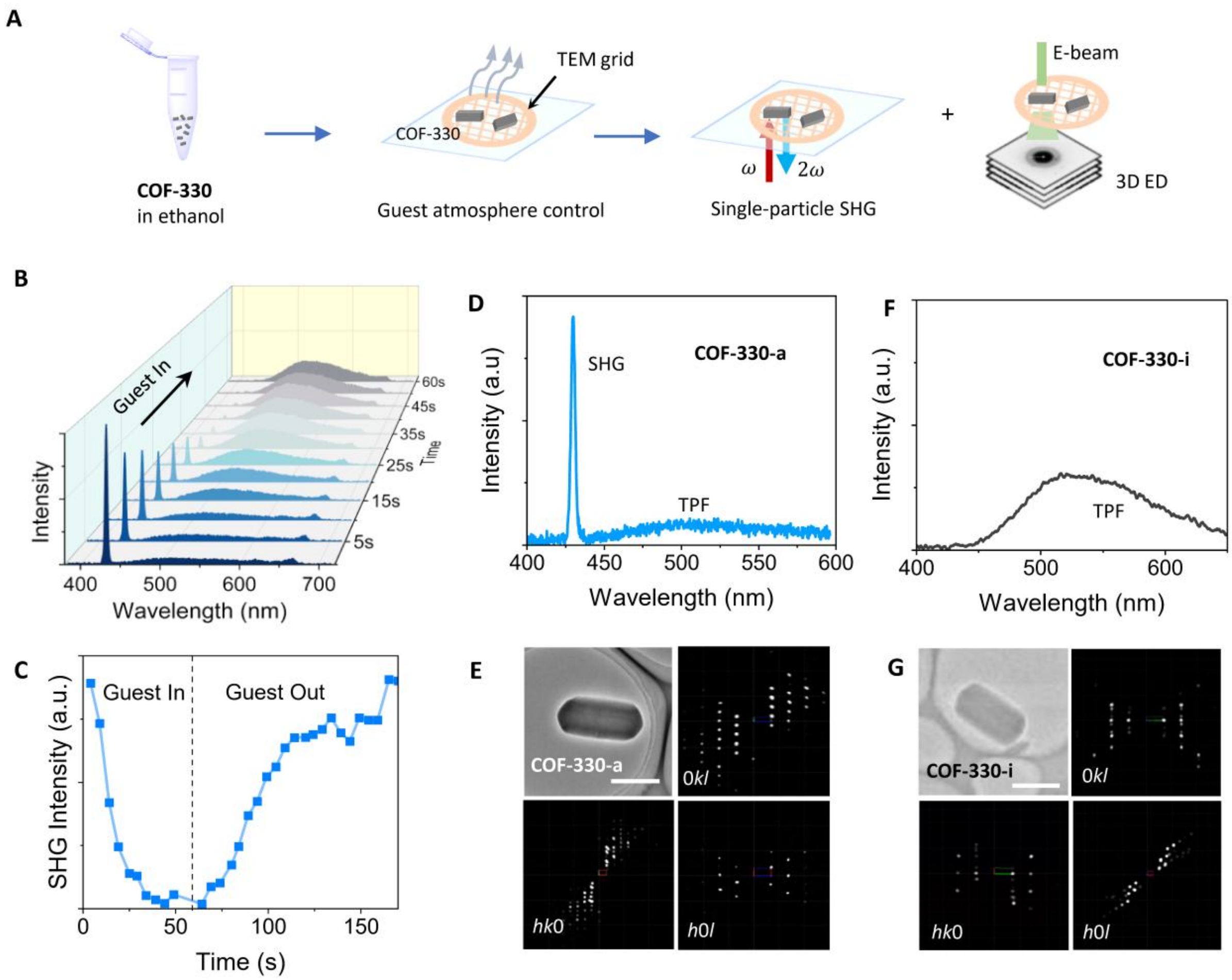

**Fig. 3. Evolution of SHG activity of individual COF-330-a.** (**A**) Schematic of 3D ED corelated single-particle SHG measurement. (**B**) Emission spectra of an individual activated microcrystal in the course of guest molecule inclusion. (**C**) Evolution of SHG intensity in the course of a guest inclusion and exclusion circle. (**D**) Emission spectrum of the individual microcrystal shown in (**E**) after guest exclusion. (**E**) TEM image and the two-dimensional slices cut from the reconstructed 3D reciprocal lattice of the microcrystal that shows SHG emission in (**D**). The reflection conditions from 3D ED data indicate the microcrystal (COF-330) adopts the space group *Pnn*2. (**F**) Emission spectrum of the individual microcrystal shown in (**F**) after guest inclusion, with SHG activity but only TPF. (**G**) TEM image and the two-dimensional slices cut from the reconstructed 3D reciprocal lattice of the microcrystal that shows no SHG emission in (**F**). The reflection conditions from 3D ED data indicate the microcrystal adopts the intermediate phase (COF-330-i) with the space group *P*2/*n*.

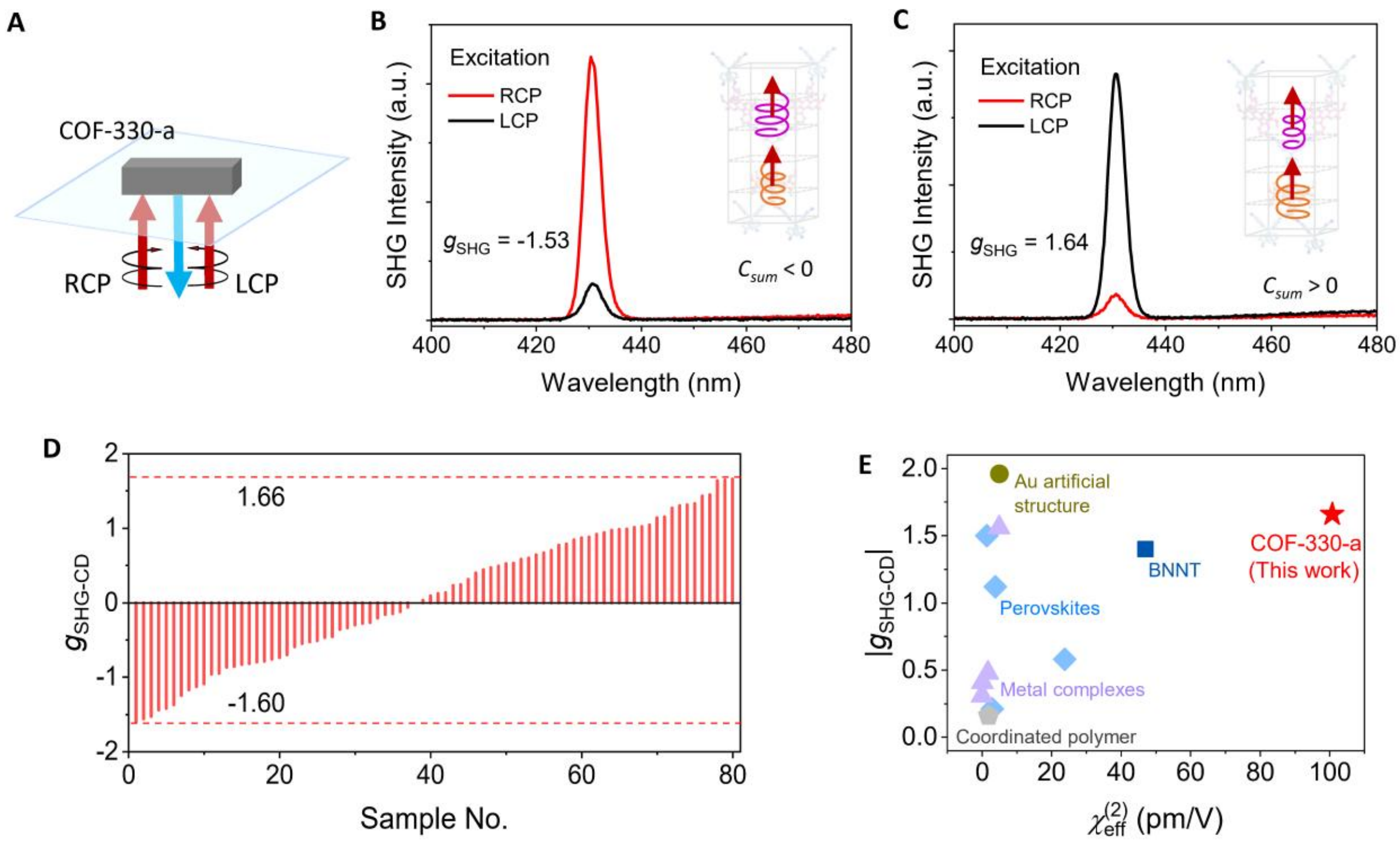

**Fig. 4. Nonlinear chiroptical response in COF-330-a. (A)** Illustration of single-particle SHG-CD measurement under RCP or LCP excitation. (**B, C**) Chiroptical SHG response of two individual COF-330-a microcrystals, indicating net chirality of opposite handedness. The microcrystal in (**B**) exhibits a stronger SHG response with LCP light excitation, and the one in (**C**) exbibits a stronger SHG response with RCP light excitation. (**D**) Plot of the measured $g_{\mathrm{SHG}}$ values of 80 individual COF-330 microcrystals. (**E**) Comparison of the chiroptical nonlinearities between COF-330-a and other state-of-the-art materials. Absolute $g_{\mathrm{SHG-CD}}$ values are plotted against the second-order nonlinear susceptibilities $\chi_{\mathrm{eff}}^{(2)}$.

# Supplementary Materials for

**Title: Chiroptical nonlinearity found in covalent organic frameworks**

**Authors**: Yile Zhang[1], Lei Wei[1], Sixue Tang[1], Yuhang Li[1], Yanhang Ma*[1,3], Yue-Biao Zhang*[1,3,4], and Xiaoqin Shen*[1,2]

**The PDF file includes:**

Materials and Methods
Supplementary Text S1-S6
Figs. S1 to S8

## Materials and Methods

COF-330 was synthesized from achiral monomers 4,4',4'',4'''-methanetetrayltetraaniline (TAM) and 4,4'-(2,3,6,7-tetramethoxyanthracene-9,10-diyl)dibenzaldehyde (AnDA), featuring an acceptor-donor-acceptor (A-D-A) π-conjugated system. The Rietveld-refined structures based on the combination of synchrotron powder X-ray diffraction (SPXRD) and three-dimensional electron diffraction (3D ED) were previously reported in ref (*17*).

The scanning electron microscopy (SEM) images of COF samples were collected on the JSM-7800F PRIME extreme-resolution field emission SEM. The samples were suspended in 1,4-dioxane or ethanol and dip-coated on the silicon sample holder or a glass cover slide for characterization under high vacuum.

The transmission electron microscopy (TEM) images and 3D electron diffraction data were collected on a TEM-JEM-F200 transmission electronic microscope equipped with a hybrid-pixel electron detector (Cheetah 1800, 512 × 512 pixels, Amsterdam Scientific Instruments). All the collected 3D ED data were visualized using the software EDT-Process, showing a resolution of about 2 Å. To collect the 3D ED data from the activated sample, the specimen was directed inserted into TEM high vacuum for data collection at room temperature. To collect the data from the COF-330 loaded with guest molecules, the specimen was frozen in liquid nitrogen before inserting into the TEM and the data were collected at around 100 K.

Power X-ray diffraction data was collected on a Bruker D8 Venture diffractometer equipped with a fine-focus Cu target X-ray tube operated at 40 W power (50 kV, 1 mA). The sample was cooled to 150 K using an Oxford Cryosystem chilled by liquid nitrogen. Bruker APEX2 software package was used for data collection.

The circular dichroism (CD) of absorption was carried out with Chirascan CD spectrometer (Applied Photophysics). The samples suspended in ethanol was drop-casted on the surface of a quartz followed by activation.

The density functional theory (DFT) calculation were carried out using the program Gaussian 09w. The ground-state dipole moment ($\mu$) and static first hyperpolarizability ($\beta$) were computed at the B3LYP/6-31G(d) level and the CAM-B3LYP/6-31G(d) level and subsequently analyzed using the Multiwfn package. The geometries of model molecules were obtained from the linkers in the XRD structures of COF-330-i and COF-330-a using Materials Studio 2023.

## Supplementary Text

### S1. Structural analysis of non-centrosymmetric dia COFs

The CIF files were analyzed by using Materials Studio 2023. For COF-330-a, the crystal structures and transformation are shown in **Fig. S1**. To reveal the universal structural features of non-centrosymmetric 3D dia COFs of point group $\bar{4}2m$ and *mm*2, we also comprehensively analyzed the skeleton geometries and the dihedral angles of the linkers of COF-303-a, LZU-310-BuOH, COF-320-op, COF-320-cp, and COF-330-a. The CIF files of these COFs were obtained from reference (*16*, *18*). As shown in **Fig. S2**, the detailed torsional angles of the linkers reveal the universal structural feature of the COFs, whereas the two upper linkers adopt right-handed chiral conformation and the lower two linkers adopt left-handed chiral conformation. Notably, for the COFs of point group *mm*2, the achiral linkers self-resolve into chiral, polar conformers of opposite handedness.

## S2. Single-particle second-harmonic generation (SHG)

The sample suspended in anhydrous ethanol or acetone was dispersed on a glass cover slide with markers and activated was activated by exclusion of the guest molecules under dry nitrogen flow in home-built sample holder. SEM image and bright field optical image of the sample were collected to identify and co-locate isolated particles for SHG measurement (**Fig. S3**).

SHG measurement was conducted on a home-built micro-spectroscope based on a Nikon Ti2-U inverted microscope, which couples with a spectrograph (Andor Sharmrock 500) and an electron-multiplying charge-coupled device (EMCCD, Newton EM971). Light beam from a Ti-sapphire femtosecond laser (Coherent Mira-HP, repletion frequency: 78 MHz, pulse width: 200 fs) was aligned into the microscope as excitation light. The laser wavelength is tunable from 800 nm to 900 nm. A half-wave plate (Thorlabs) was placed in the optical path before the beam enters the microscope. Polarization angle of the excitation light was controlled by rotating the HWP with stepper motor. The linearly polarized laser was then focused on samples by a 60x infinite conjugate objective. The generated signals reflected from samples is collected by the same objective and transported into the spectrograph. A quarter-wave plate (Thorlabs) was placed behind the half-wave plate for generating the left-circulated polarization and right-circulated polarization excitation light for the measurement of SHG circular dichroism (SHG-CD).

For corelating single-particle SHG with 3D ED, the sample was first dispersed on a TEM copper grid which was then placed on a glass cover slide for the SHG measurement. After the nonlinear optical measurement, the sample on the copper grid was used for 3D ED measurement.

## S3. Resonance enhancement of SHG

According to density matrix formulation of quantum mechanics, the band structure could be simplified into a two-level system which are correspondingly denoted as ground states $\psi_g\rangle$ and excited states $\psi_{exc}\rangle$. In the case of that there exists a virtual intermediate state $\psi_{vi}\rangle$ possessing a band gap resonant with the energy of pump photons. We can then derive the second-order susceptibility:

$$\chi_{ijk}^{(2)}(-2\omega;\omega,\omega) \propto \frac{\langle\psi_g|\hat{V}_i^{2\omega}|\psi_{exc}\rangle}{E_{exc}-E_g-2h\omega-i\Gamma_{exc\to g}} \cdot \frac{\langle\psi_{exc}|\hat{V}_j^{\omega}|\psi_{vi}\rangle\langle\psi_{vi}|\hat{V}_k^{\omega}|\psi_g\rangle}{(E_{vi}-E_g-h\omega-i\Gamma_{vi\to g})} \quad (1)$$

Here, the $\hat{V}$ operator represents the interactions happened during SHG process, including electric dipole interaction, magnetic dipole interaction and electric quadrupole interactions. $\Gamma_{exc\to g}$ and $\Gamma_{vi\to g}$ are the decay rate of the real excited state and the virtual state to the ground state, respectively. Given that $\Gamma_{exc\to g} \ll \Gamma_{vi\to g}$ as the lifetime of real excited states is much longer than the virtual state, $\chi_{ijk}^{(2)}(-2\omega;\omega,\omega)$ mainly depends on $E_{exc} - E_g - 2h\omega$ in the denominator of the first term. Therefore, $\chi_{ijk}^{(2)}$ will increase sharply as $(E_{exc} - E_g) \to 2h\omega$. It means that SHG response will be resonantly enhanced when the energy of pump photons is close enough to half of the band gap energy. As shown in **Fig. S4**, the strong SHG response of (quasi)racemic COF-330 under exaction at about 890 nm is owing the resonance enhancement of SHG.

## S4. Evaluation of laser damage threshold

In the experiment, the power the femtosecond laser is adjusted by a set of neutral density filter with different optical density. The laser beam was focused by a 60× objective (NA=0.65) to a spot with diameter $d \approx 0.6$ μm. SHG response of isolated COF-330 particles was collected under excitation by the laser at 860 nm with the average power in the range of 0.4 mW to 4 mW. The peak power intensities were calculated by $I = \frac{P_{\text{peak}}}{A}$, where $P_{\text{peak}}$ is the peak power of the laser pulse, A is the area of the effective laser spot. The measured SHG intensities and the corresponding peak power densities were plotted on a log - log scale coordinate (**Fig. S5**). The SHG signal exhibited a quadratic dependence on the peak power density bellow about 30 $GW/cm^2$. Above 30 $GW/cm^2$, the dependency between SHG and pump deviated from quadratic. The SHG intensity sharply dropped at about 100 $GW/cm^2$, and damage of isolated COF was observed through the optical microscope. These results suggested that (quasi)racemic COF-330 exhibit a high laser damage threshold of about 30 to 90G $GW/cm^2$.

## S5. Evaluation of second-order nonlinear susceptibility ($\chi_{eff}^{(2)}$)

The $\chi_{eff}^{(2)}$ of (quasi)racemic COF-330 was calculated by using urea crystals ($\chi^{(2)}$ = 2.8 pm/V) of similar size as a reference under the same SHG measurement condition. SHG intensity of a certain sample can be written in the equation below:

$$I_{2\omega}(L) = \frac{2\omega^2}{c^3 n_\omega^2 n_{2\omega}\epsilon_0}\left|d_{eff}\right|^2 I_\omega^2 L^2 \frac{\sin^2\left(\frac{\Delta kL}{2}\right)}{\left(\frac{\Delta kL}{2}\right)^2} \qquad (2)$$

Where $n_\omega$ and $n_{2\omega}$ are the refractive index of the samples at the SH and pump wavelength, respectively. $L$ is the thickness of the sample, which is of about the excitation wavelength. $\frac{\sin^2\left(\frac{\Delta kL}{2}\right)}{\left(\frac{\Delta kL}{2}\right)^2}$ is the phase condition factor for thick samples. Because that $L_{urea}$, $L_{COF} \sim \lambda$, $\Delta kL \rightarrow 0$, the phase condition factor can be approximately regard as 1. The $\chi_{eff}^{(2)}$ of COF-330-a can be obtained as

$$\chi_{eff}^{(2)}(\mathrm{COF}) = \chi_{eff}^{(2)}(urea)\sqrt{\frac{I_{2\omega}^{COF}\left(n_\omega^{COF}\right)^2 n_{2\omega}^{COF}\left(I_\omega^{urea}\right)^2 L_{urea}^2}{I_{2\omega}^{urea}\left(n_\omega^{urea}\right)^2 n_{2\omega}^{urea}\left(I_\omega^{COF}\right)^2 L_{COF}^2}} \quad (3)$$

The refractive indices of COF-330-a were estimated to be 1.61 and 1.67 at the SH and pump wavelength, respectively, by measuring the Brewster angles. By applying the measured SHG intensities into the expression (3), $\chi_{eff}^{(2)}$ of the (quasi)racemic COF-330 was obtained to be ~ 62 pm/V at 860 nm and ~101 pm/V at 890 nm.

## S6. Monitoring the NLO responses of individual microcrystals

The experiment was carried out with a customized sample holder on the Nikon Ti2-U inverted microscope. The sample holder was connected with an atmosphere control system (**Fig. S7**) which consists of two mass flow meters and a reservoir of anhydrous ethanol. One mass flow meter serves for a pure nitrogen gas route, the other serves for the working gas route which contains anhydrous ethanol vapor in a nitrogen flow. The two routes of gas flow were then merged into one and was introduced to the customized sample holder. The flow rate ratio between the two gas routes is controlled by two mass flow meters to regulate the relative gas ratio of two routes. Single-particle SHG spectrum was continuously captured at 5 seconds intervals upon the gas flow entering into the sample holder at an optimized flow rate ratio ($v_{EtOH}: v_{N_2} = 40:60$). The SHG response progressively diminished to zero in about 60 seconds. After that, pure nitrogen gas flow was introduced into the sample holder, the spectrum was continued to be captured until SHG response was fully recovered in about 110 seconds.

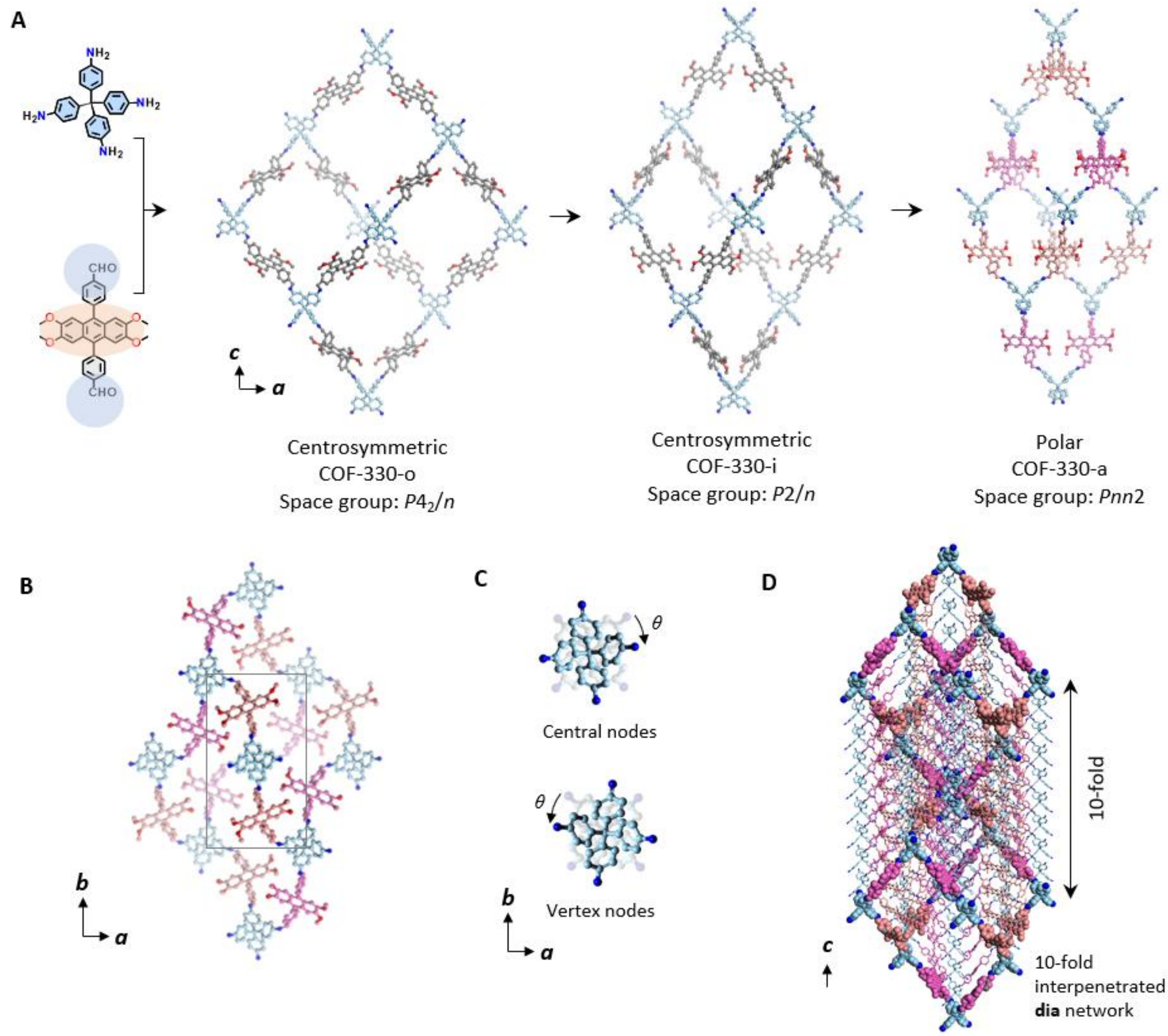

**Fig. S1. Crystal structure of COF-330.** (**A**) Formation and structural transformation of COF-330-a. COF-330-o was synthesized from achiral monomers 4,4',4'',4'''-methanetetrayltetraaniline (TAM) and 4,4'-(2,3,6,7-tetramethoxyanthracene-9,10-diyl)dibenzaldehyde (AnDA) which features an acceptor-donor-acceptor (A-D-A) π-conjugated system. The framework transforms from the centrosymmetric COF-330-o to COF-330-i and to COF-330-a**.** (**B**) **dia** cage structure COF-330-a viewed along crystalline *c* axis. The grey rectangle indicates the lattice. (**C**) Highlight of node rotations in COF-330-a compared to COF-330-i. The central and vertex nodes of unit cell rotated clockwise and counterclockwise of 39°, respectively. (**D**) Structures of a 10-fold interpenetrated **dia** network of COF-330-a, whereas the first and tenth **dia** cages are highlighted.

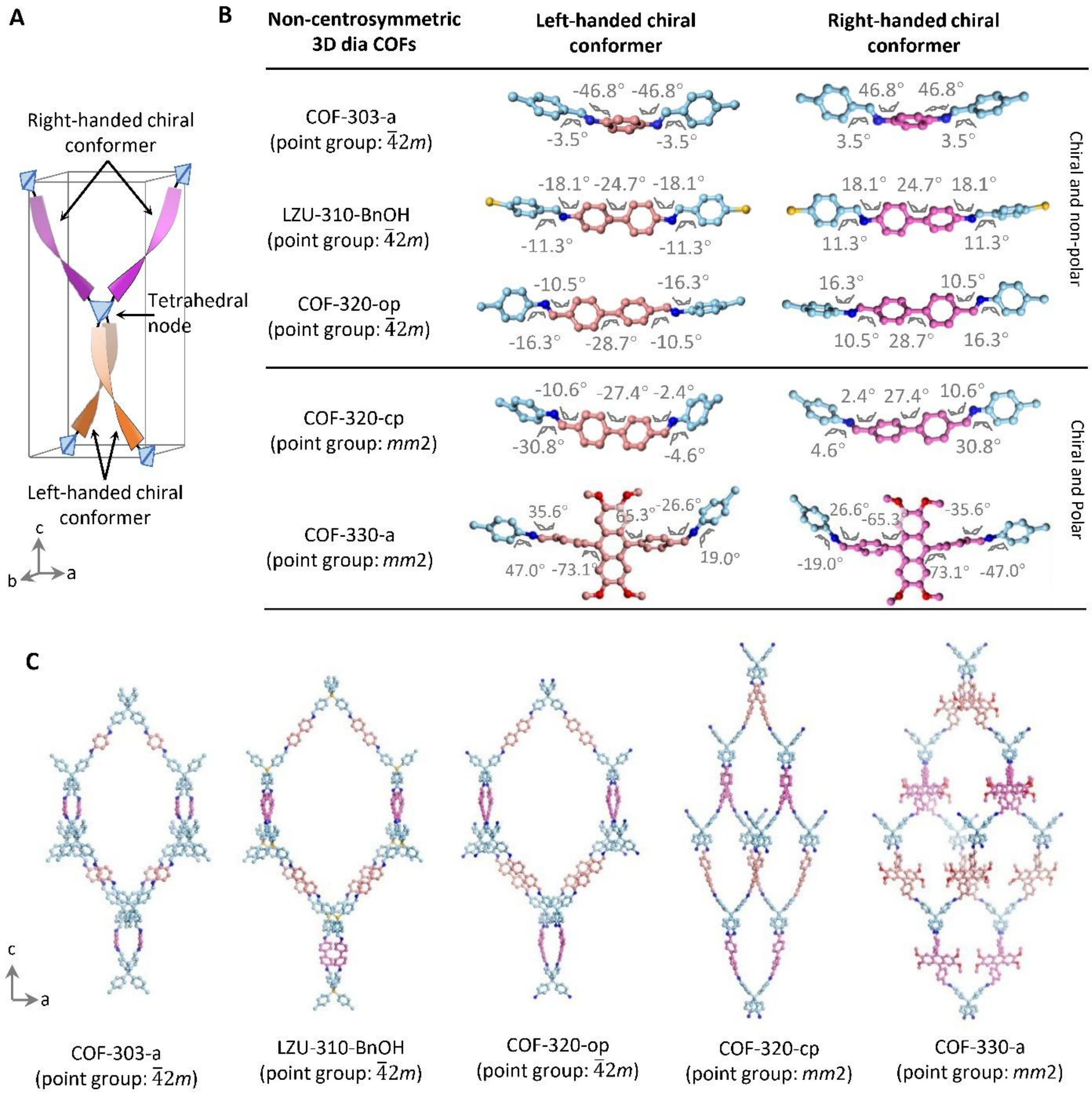

**Fig. S2. Universal structural features of non-centrosymmetric 3D dia COFs of point group $\bar{4}2m$ and $mm2$.** (**A**) Schematic of the universal structural feature of the four-armed repeating unit of the COFs, whereas the two upper linkers adopt right-handed chiral conformation and the lower two linkers adopt left-handed chiral conformation. **(B)** Torsional angle analysis of the linkers of COF-303-a, LZU-310-BuOH, COF-320-op, COF-320-cp and COF-330-a. Notably, for the COFs of point group *mm*2, the achiral linkers self-resolve into chiral, polar conformers of opposite handedness. **(C) dia** cage structures of COF-303-a, LZU-310-BuOH, COF-320-op and COF-320-cp. For COF-330-a and COF-330-cp, each forms a polar order along *c* axis.

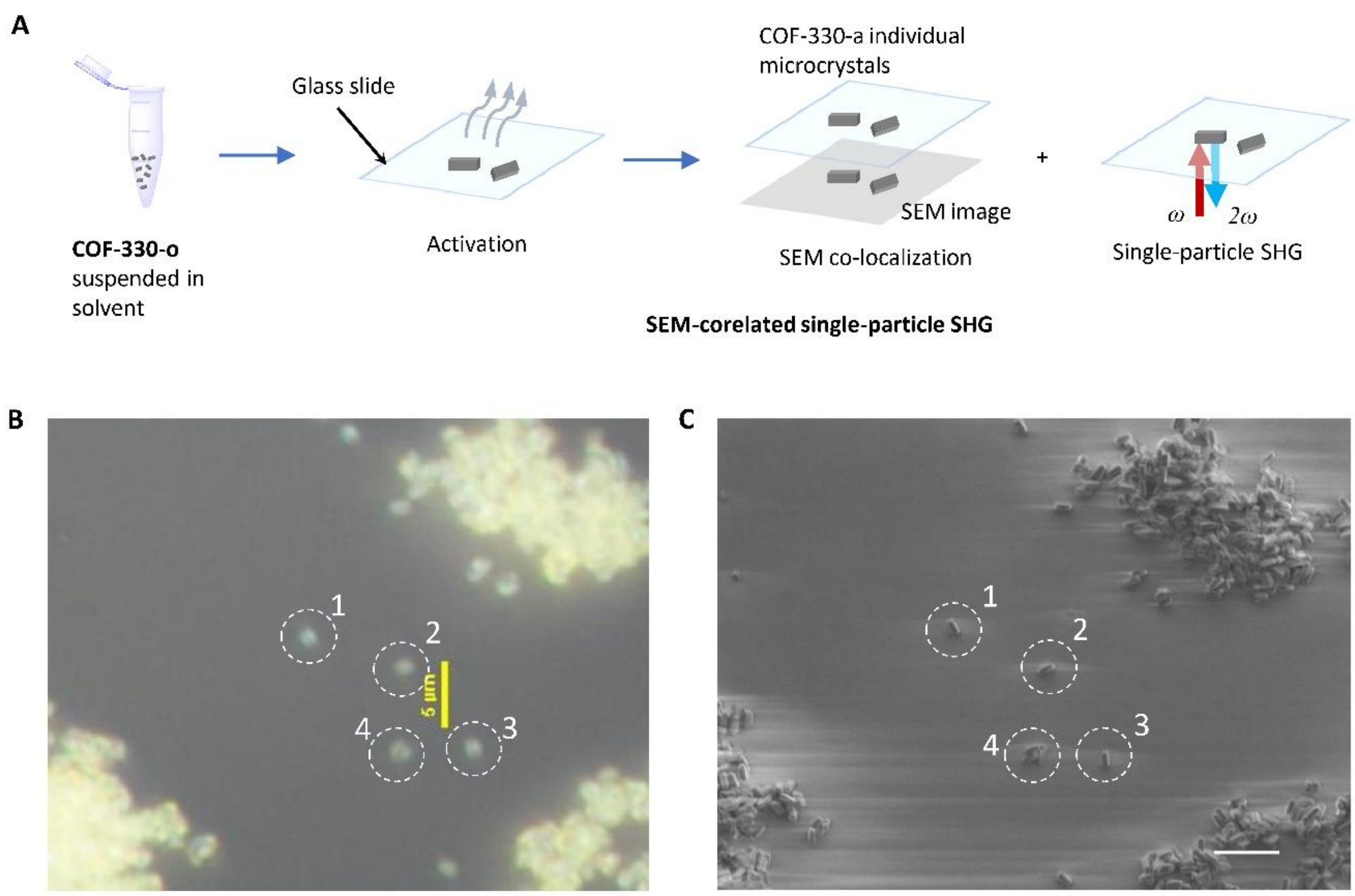

**Fig. S3. Single-particle SHG measurement.** **(A)** Schematic of SEM-correlated SHG. Image and bright field optical image of the sample were collected to identify individual microrystals for SHG measurement under a back-reflection configuration. (**B**) A representative optical microscopy image of the sample dispersed on a glass slide. The dashed circles show four isolated image spots from the sample. **(C)** The corresponding SEM image of the sample in (A). No.1-3 are identified as single particles. No. 4 was identified to be a random aggregate of three microcrystals and was excluded for SHG analysis. Scale bar: 5 μm.

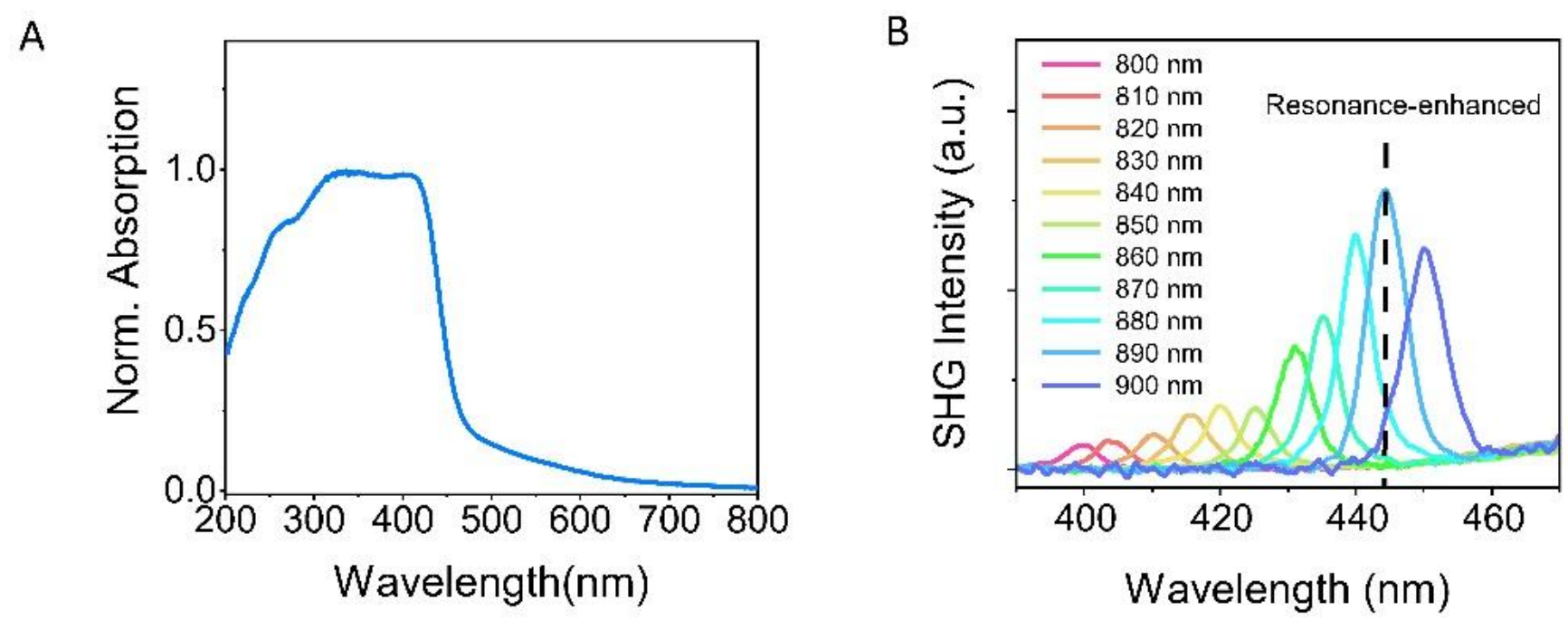

**Fig. S4. Resonance-enhanced SHG.** (**A**) Normalized absorption of activated COF-330-a. It shows the band gap energy to approximate the photon energy at 440 to 450 nm. (**B**) Dependency of SHG intensity to the pump wavelength from 800 nm to 900 nm (same as Fig 2C in the main text). Maximum SHG response appeared at 450 nm, under excitation wavelength of 880 nm to 900 nm, consistent with the band gap energy indicated in (**A**). There results indicate significant resonance enhanced of SHG at ~445 nm in activated COF-330-a. Self-absorption of SHG is not negligible at shorter wavelength (400 nm to 445 nm).

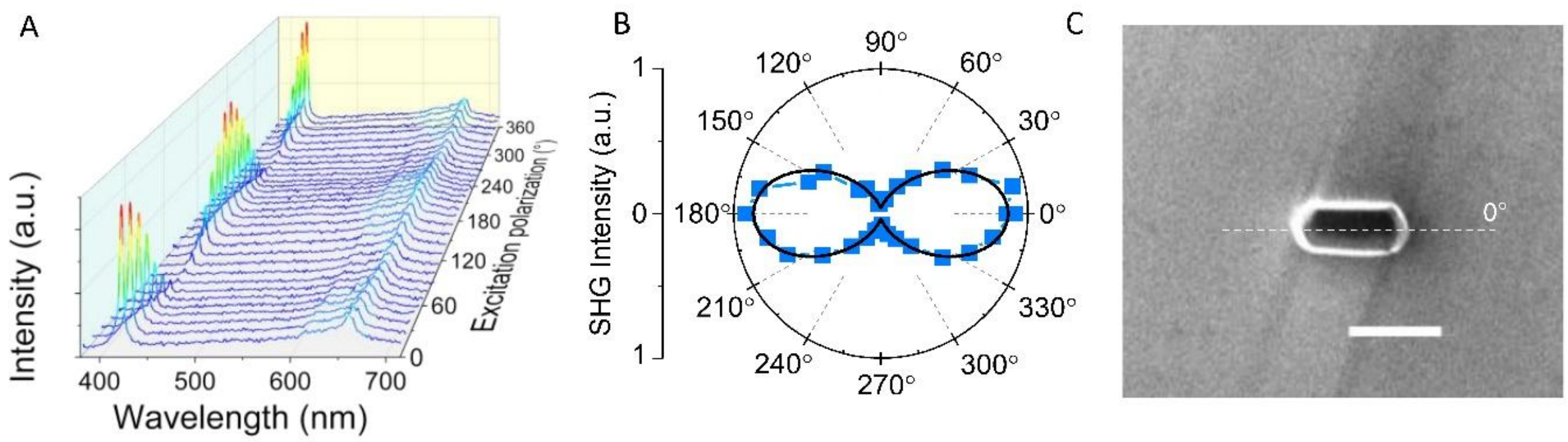

**Fig. S6. Polarization-dependent SHG.** (**A**) The representative SHG spectra under excitation at different polarization angle. (**B**) The polar plot of the SHG intensity (Same as Fig 2E in the main text). (**C**) SEM image of an individual activated COF-330 for the measurement. The polarization angle along the longitudinal axis of the sample was set zero.

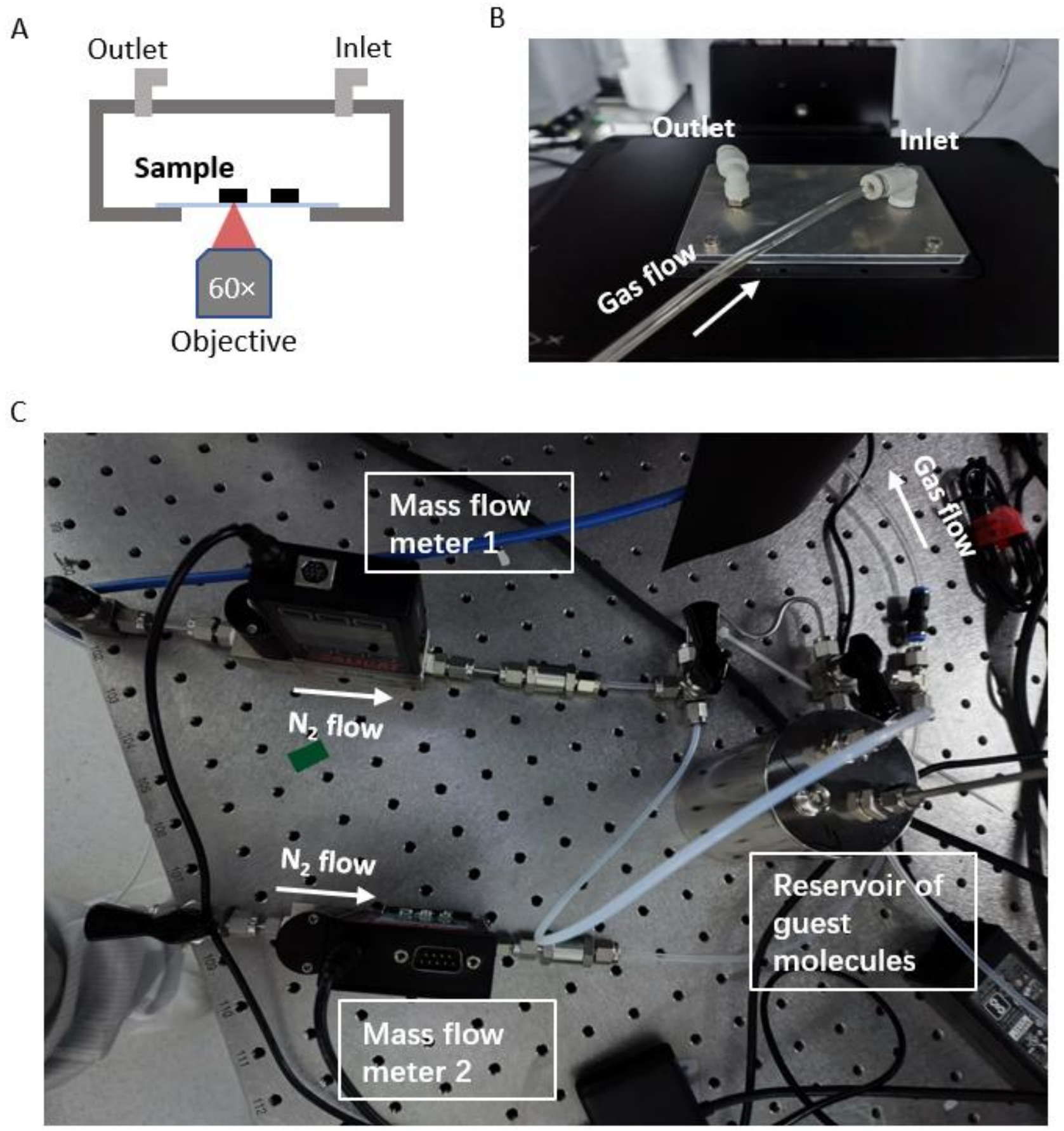

**Fig. S7. Customized setup for single-particle measurement and real time monitoring of SHG.** (**A**) Illustration of the customized sample holder in the working condition. (**B**) Image of the sample holder mounted on the stage of Nikon Ti2-U microscope. (**C**) Image of the atmosphere control system. It consists of two mass flow meters and a reservoir of anhydrous ethanol. One mass flow meter serves for a pure nitrogen gas route, the other serves for the working gas route which contains anhydrous ethanol vapor in a nitrogen flow. The two routes of gas flow were then merged into one and was introduced to the customized sample holder (**B**).

A B

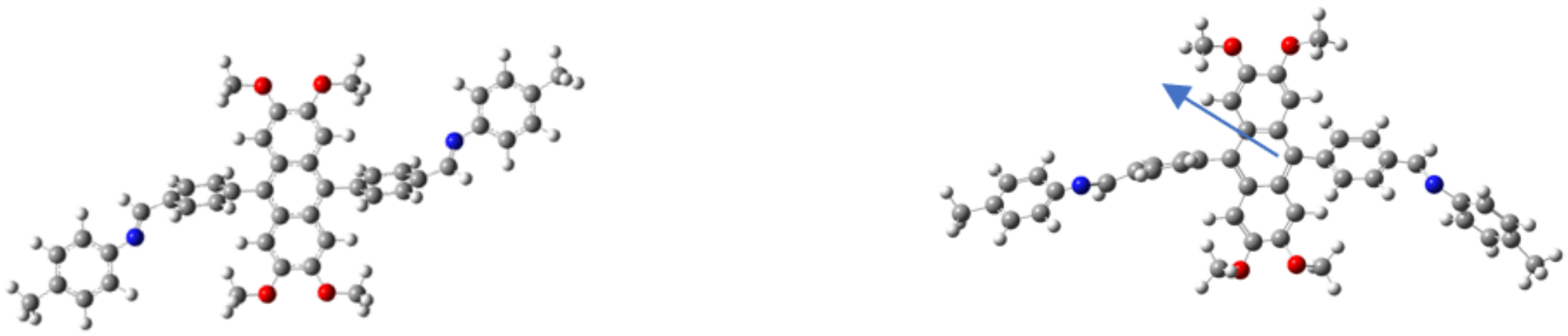

Model structure for linker of COF-330-i — Model structure for linker of COF-330-a

$\mu = 0.068$ D; $\beta = 1.02 \times 10^{-30}$ esu — $\mu = 4.906$ D; $\beta = 61.98 \times 10^{-30}$ esu

**Fig. S8. DFT calculation results of ground state dipole moments ($\mu$) and first-order molecular hyperpolarizabilities ($\beta$).** (**A**) and (**B**) show the model molecules for the linker conformation of COF-330-i and COF-330-a, respectively. The blue arrow indicates the molecular dipole moment $\mu$. The computed results of $\mu$ and $\beta$ at the CAM-B3LYP/6-31G(d) level listed below the model molecules. The results indicate that the asymmetric twisting of the large π-electron conjugated linkers enable a substantial electronic dipole moment and a large molecular hyperpolarizability. In contrast, the model molecules for linkers of centrosymmetric COF-330-i have negligible $\mu$ and $\beta$.